\documentstyle[twoside,graphicx,verbatim,url]{article}

\newcommand{\pr}[1]{\left(#1\right)}
\newcommand{\cro}[1]{\left[#1\right]}
\newcommand{\avg}[1]{\langle{#1}\rangle}
\newcommand{\req}[1]{(\ref{#1})}
\newcommand{\be}{\begin{equation}}
\newcommand{\ee}{\end{equation}}
\newcommand{\bea}{\begin{eqnarray}}
\newcommand{\eea}{\end{eqnarray}}
\newcommand{\erf}{\mbox{erf}}
\newcommand{\dd}{\mbox{d}}

\catcode`\@=11
\long\def\@makefntext#1{
\protect\noindent \hbox to 3.2pt {\hskip-.9pt  
$^{{\eightrm\@thefnmark}}$\hfil}#1\hfill}		

\def\thefootnote{\fnsymbol{footnote}}
\def\@makefnmark{\hbox to 0pt{$^{\@thefnmark}$\hss}}	
	
\def\ps@myheadings{\let\@mkboth\@gobbletwo
\def\@oddhead{\hbox{}
\rightmark\hfil\eightrm\thepage}   
\def\@oddfoot{}\def\@evenhead{\eightrm\thepage\hfil
\leftmark\hbox{}}\def\@evenfoot{}
\def\sectionmark##1{}\def\subsectionmark##1{}}

\oddsidemargin=\evensidemargin
\renewcommand{\thefootnote}{\fnsymbol{footnote}}

\newcounter{sectionc}\newcounter{subsectionc}\newcounter{subsubsectionc}
\renewcommand{\section}[1] {\vspace{12pt}\addtocounter{sectionc}{1} 
\setcounter{subsectionc}{0}\setcounter{subsubsectionc}{0}\noindent 
	{\tenbf\thesectionc. #1}\par\vspace{5pt}}
\renewcommand{\subsection}[1] {\vspace{12pt}\addtocounter{subsectionc}{1} 
	\setcounter{subsubsectionc}{0}\noindent 
	{\bf\thesectionc.\thesubsectionc. {\kern1pt \bfit #1}}\par\vspace{5pt}}
\renewcommand{\subsubsection}[1] {\vspace{12pt}\addtocounter{subsubsectionc}{1}
	\noindent{\tenrm\thesectionc.\thesubsectionc.\thesubsubsectionc.
	{\kern1pt \tenit #1}}\par\vspace{5pt}}

\newcounter{appendixc}
\newcounter{subappendixc}[appendixc]
\newcounter{subsubappendixc}[subappendixc]
\renewcommand{\thesubappendixc}{\Alph{appendixc}.\arabic{subappendixc}}
\renewcommand{\thesubsubappendixc}
	{\Alph{appendixc}.\arabic{subappendixc}.\arabic{subsubappendixc}}

\renewcommand{\appendix}[1] {\vspace{12pt}
        \refstepcounter{appendixc}
        \setcounter{figure}{0}
        \setcounter{table}{0}
        \setcounter{lemma}{0}
        \setcounter{theorem}{0}
        \setcounter{corollary}{0}
        \setcounter{definition}{0}
        \setcounter{equation}{0}
        \renewcommand{\thefigure}{\Alph{appendixc}.\arabic{figure}}
        \renewcommand{\thetable}{\Alph{appendixc}.\arabic{table}}
        \renewcommand{\theappendixc}{\Alph{appendixc}}
        \renewcommand{\thelemma}{\Alph{appendixc}.\arabic{lemma}}
        \renewcommand{\thetheorem}{\Alph{appendixc}.\arabic{theorem}}
        \renewcommand{\thedefinition}{\Alph{appendixc}.\arabic{definition}}
        \renewcommand{\thecorollary}{\Alph{appendixc}.\arabic{corollary}}
        \renewcommand{\theequation}{\Alph{appendixc}.\arabic{equation}}
        \noindent{\tenbf Appendix \theappendixc #1}\par\vspace{5pt}}
\newcommand{\subappendix}[1] {\vspace{12pt}
        \refstepcounter{subappendixc}
        \noindent{\bf Appendix \thesubappendixc. {\kern1pt \bfit #1}}
	\par\vspace{5pt}}
\newcommand{\subsubappendix}[1] {\vspace{12pt}
        \refstepcounter{subsubappendixc}
        \noindent{\rm Appendix \thesubsubappendixc. {\kern1pt \tenit #1}}
	\par\vspace{5pt}}

\newcommand{\textlineskip}{\baselineskip=13pt}
\newcommand{\smalllineskip}{\baselineskip=10pt}

\def\eightcirc{
\begin{picture}(0,0)
\put(4.4,1.8){\circle{6.5}}
\end{picture}}
\def\eightcopyright{\eightcirc\kern2.7pt\hbox{\eightrm c}} 

\newcommand{\copyrightheading}[1]
	{\vspace*{-2.5cm}\smalllineskip{\flushleft
	{\footnotesize International Journal of Reliability, 
	 Quality and Safety Engineering #1}\\
	{\footnotesize $\eightcopyright$\, World Scientific Publishing
	 Company}\\
	 }}

\newcommand{\publisher}[2]{{\begin{center}\footnotesize\smalllineskip 
	Received #1\\
	Revised #2
	\end{center}
	}}

\def\abstracts#1#2#3{{
	\centering{\begin{minipage}{4.5in}\baselineskip=10pt\footnotesize
	\parindent=0pt #1\par 
	\parindent=15pt #2\par
	\parindent=15pt #3
	\end{minipage}}\par}} 

\def\keywords#1{{
	\centering{\begin{minipage}{4.5in}\baselineskip=10pt\footnotesize
	{\footnotesize\it Keywords}\/: #1
	\end{minipage}}\par}}

\renewenvironment{thebibliography}[1]
        {\frenchspacing
	 \ninerm\baselineskip=11pt
         \begin{list}{\arabic{enumi}.}
        {\usecounter{enumi}\setlength{\parsep}{0pt}     
	 \setlength{\leftmargin 12.7pt}{\rightmargin 0pt} 
         \setlength{\itemsep}{0pt} \settowidth
	{\labelwidth}{#1.}\sloppy}}{\end{list}}

\newcounter{itemlistc}
\newcounter{romanlistc}
\newcounter{alphlistc}
\newcounter{arabiclistc}

\newcommand{\fcaption}[1]{
        \refstepcounter{figure}
        \setbox\@tempboxa = \hbox{\footnotesize Fig.~\thefigure. #1}
        \ifdim \wd\@tempboxa > 5in
           {\begin{center}
        \parbox{5in}{\footnotesize\smalllineskip Fig.~\thefigure. #1}
            \end{center}}
        \else
             {\begin{center}
             {\footnotesize Fig.~\thefigure. #1}
              \end{center}}
        \fi}

\newcommand{\tcaption}[1]{
        \refstepcounter{table}
        \setbox\@tempboxa = \hbox{\footnotesize Table~\thetable. #1}
        \ifdim \wd\@tempboxa > 5in
           {\begin{center}
        \parbox{5in}{\footnotesize\smalllineskip Table~\thetable. #1}
            \end{center}}
        \else
             {\begin{center}
             {\footnotesize Table~\thetable. #1}
              \end{center}}
        \fi}

\def\@citex[#1]#2{\if@filesw\immediate\write\@auxout
	{\string\citation{#2}}\fi
\def\@citea{}\@cite{\@for\@citeb:=#2\do
	{\@citea\def\@citea{,}\@ifundefined
	{b@\@citeb}{{\bf ?}\@warning
	{Citation `\@citeb' on page \thepage \space undefined}}
	{\csname b@\@citeb\endcsname}}}{#1}}

\newif\if@cghi
\def\cite{\@cghitrue\@ifnextchar [{\@tempswatrue
	\@citex}{\@tempswafalse\@citex[]}}
\def\citelow{\@cghifalse\@ifnextchar [{\@tempswatrue
	\@citex}{\@tempswafalse\@citex[]}}
\def\@cite#1#2{{$\null^{#1}$\if@tempswa\typeout
	{IJCGA warning: optional citation argument 
	ignored: `#2'} \fi}}

\def\pmb#1{\setbox0=\hbox{#1}
	\kern-.025em\copy0\kern-\wd0
	\kern.05em\copy0\kern-\wd0
	\kern-.025em\raise.0433em\box0}

\def\fnt#1#2{\footnotetext{\kern-.3em
	{$^{\mbox{\scriptsize #1}}$}{#2}}}

\def\fpage#1{\begingroup
\voffset=.3in
\thispagestyle{empty}\begin{table}[b]\centerline{\footnotesize #1}
	\end{table}\endgroup}

\def\runninghead#1#2{\pagestyle{myheadings}
\markboth{{\protect\footnotesize\it{\quad #1}}\hfill}
{\hfill{\protect\footnotesize\it{#2\quad}}}}
\font\tenrm=cmr10
\font\tenit=cmti10 
\font\tenbf=cmbx10
\font\bfit=cmbxti10 at 10pt
\font\ninerm=cmr9

\font\eightrm=cmr8

\def\qed{\hbox{${\vcenter{\vbox{			
   \hrule height 0.4pt\hbox{\vrule width 0.4pt height 6pt
   \kern5pt\vrule width 0.4pt}\hrule height 0.4pt}}}$}}

\renewcommand{\thefootnote}{\fnsymbol{footnote}}	

\def\bsc{{\sc a\kern-6.4pt\sc a\kern-6.4pt\sc a}}  	
\def\bflatex{\bf L\kern-.30em\raise.3ex\hbox{\bsc}\kern-.14em 
T\kern-.1667em\lower.7ex\hbox{E}\kern-.125em X} 

\begin{document}

\runninghead{MICROSCOPIC MODEL OF SOFTWARE BUG DYNAMICS:}
{International Journal of Reliability, Quality and Safety Engineering}

\normalsize\textlineskip
\thispagestyle{empty}
\setcounter{page}{1}

\copyrightheading{}			

\vspace*{0.88truein}

\fpage{1}
\centerline{\bf MICROSCOPIC MODEL OF SOFTWARE BUG DYNAMICS:}
\vspace*{0.035truein}
\centerline{\bf  CLOSED SOURCE VERSUS OPEN SOURCE}
\vspace*{0.37truein}
\centerline{\footnotesize DAMIEN CHALLET\footnote{challet@maths.ox.ac.uk}}
\vspace*{0.015truein}
\centerline{\footnotesize\it Nomura Centre for Quantitative Finance, Mathematical Institute, Oxford University, 24--29 St-Gile's}
\baselineskip=10pt
\centerline{\footnotesize\it Oxford OX1 3LB, United Kingdom}
\vspace*{10pt}
\centerline{\normalsize and}
\vspace*{10pt}
\centerline{\footnotesize YANN LE DU\footnote{yann@thphys.ox.ac.uk}}
\vspace*{0.015truein}
\centerline{\footnotesize\it The Rudolf Peierls Centre for Theoretical Physics, Oxford University, 1--3 Keble Road}
\baselineskip=10pt
\centerline{\footnotesize\it Oxford, OX1 3NP, United Kingdom}
\vspace*{0.225truein}
\publisher{(received date)}{(revised date)}

\vspace*{0.21truein}
\abstracts{We introduce a simple microscopic description of software bug dynamics where
  users, programmers and a maintainer interact through a given program, with a particular emphasis on bug creation, detection and fixing.
  When the program is written from scratch, the first phase of
  development is characterized by a fast decline of the number of
  bugs, followed by a slow phase where most bugs have been fixed,
  hence, are hard to find. Releasing immediately bug fixes speeds up
  the debugging process, which substantiates bazaar open-source
  methodology. We provide a mathematical analysis that supports our numerical simulations. Finally, we apply our model to Linux history and determine the existence of a lower bound 
  to the quality of its  programmers.
}{}{}

\vspace*{10pt}


\vspace*{1pt}\textlineskip	
\vspace*{-0.5pt}
\noindent

\pagebreak

\textheight=7.8truein
\setcounter{footnote}{0}
\renewcommand{\thefootnote}{\alph{footnote}}

\section{Introduction}
\noindent
The importance of reliable software is obvious nowadays, as computers
control an growing part of our life. At the same time the complexity
of software is ever increasing.  A particularly important problem is
the management of software projects so as to minimize development cost
or release software on time, while ensuring the quality of software.
Appropriate methodologies of programming and team synchronization are
designed to avoid errors in the first place\cite{SoftwareEng,Agile}, but it is common wisdom that software development goes through a cycle of trial and errors and that it is very hard to estimate the overal quality of software and the time needed to deliver a well-tested product. Theoretical approaches of software reliability are done via so called reliability growth models, which focus on how much errors a program contains, that is, the failure rate, or the time between two failures, with a particular emphasis on predicting these quantities (for a review, see Lyu's book\cite{Lyu}, Chapter 3). There is a hierarchy of approaches: early studies are of macroscopic nature: one measures the evolution of a quantity that depends on the errors in a given program, such as the time to next failure. In order to make useful predictions, one uses fitting functions, also known as models. A more recent approach is mesoscopic, that is, splits a program in several modules, each having their own propensity to fail. Remarkably, the evolution of the global failure rate is universal, and the latter decreases as power-law, that is, very slowly\cite{Bishop,RossBall}.

Our work supplements current vast litterature on the topic\cite{web} by describing what happens when a bug is fixed. Indeed, almost all the reliability growth models assume that a bug is fixed as soon as it is detected and that doing so nothing else is broken. We propose to remedy that by using a simplified microscopic framework where the fundamental unit is the software bug, not code units modules or even submodules. Doing so, one may indentify fundamental microscopic laws, and above all, the crucial ingredients of software reliability: in this paper, we not interested in {\em predicting} reliability, but in understanding on what microscopic processes it depends. Starting at the microscopic level and being able to compute analytically what macroscopic properties emerge brings much insights: a large part the success of modern Physics come from the development of microscopic models and mathematical methods for dealing with a great many of elementary units. What is meant by {\em model} is not a fitting function, that is, an arbitrary a priori relationship between measurables that fits more or less convincingly data, but how elementary units are born, interact with each other or with external constraints, and die. Such microscopic models are usually much simplified, but capture the {\em essence} of a given situation because they contain the most fundamental interactions that give rise to relevant phenomenology. For instance, a atom of iron contains 26 protons, 26 neutrons and 26 electrons, each in interaction with each other and with those of the neighbouring atoms; in order to understand how microscopic interactions lead to magnetism, a collective phenomenon, it is enough to replace each atom with one binary variable and assume a nearest-neighbour interaction. This is the path that we propose to follow in this article. Our aim is to keep the most important {\em microscopic} interactions in a very simple and extendable framework that allows for more realistic ingredients to be added if need be, but that should reproduce qualitatively important properties of software quality evolution. The number of parameters must be very small, otherwise one is very likely to be unable to understand the relative importance of each ingredient. 

Our model shows that software projects
can converge to a bug-free state even with imperfect programmers and maintainers. However, if there is any number of users that fill bug reports on imaginary problems and if the programmers modify the code without double-checking, the bug-free state is not a fixed point of the dynamics. Our model is also able to explain why programs whose source code is openly available, the so-called open-source software (OSS)\cite{fsf}, are able to reach a high quality, despite the fact that its programmers are sometimes less skilled than those working for closed-source software (CSS) companies. OSS has become quite relevant because of the rise of successful open-source project such as Linux\cite{linux} or Apache.\cite{apache}
There are broadly speaking two types of open source projects, often
called bazaar and cathedral, as put by Raymond.\cite{ESR}  Bazaar
projects such as Linux release new versions "early and often", and welcome contributions by everybody, while cathedral projects release new versions at a lower pace, and are crafted by a smaller group of programmers. In that respect cathedral OSS stands between CSS and bazaar OSS.

\section{Definition of the model}
\noindent
Our approach is to create a minimal model, that does not describe in all its sublety how software testing is done, but the microscopic processes of bug discovery and elimination, which are remarkably absent from current literature, where it is assumed that a bug is corrected immediately when discovered. Being minimal, this model allows for easy addition of relevant extensions; we shall discuss some of them in the Outlook Section.

In our model, a program is defined as a collection of $L$ parts $i=1,\cdots,L$, or modules; a part provides a basic functionality such as file loading. Each part has $M$ subparts. The total number of subparts $LM$ will be referred to as the size $S$ of the project. We shall make the assumption of independence, which means that if a given (sub)part is buggy, the other (sub)parts are not affected; this is clearly not the case in real life, as bug influence propagate\cite{SoftLinks} on the functional/object dependence network\cite{Sole} (see also \cite{Lyu}, Chapter 13, pp 538). Another assumption is that all the parts have the same number of sub-parts; this is a less important assumption that can be easily remedied by assuming a probability distribution function for the number of subparts of each part $i$, denoted by $M_i$, which would increase the number of parameters of the model.

Let us introduce some important notations that characterize the state of the program at time $t$. Subpart $j$ of part
  $i$ ($j=1,\cdots,M$) is either bug free --- in which case its state is denoted by $s_{i,j}(t)=0$ ---, or buggy ($s_{i,j}(t)=1$). A 
feature request is considered as a bug.  At time $t$, part $i$ has $b_i(t)=\sum_{j=1}^Ms_{i,j}(t)$ bugs, and the total number of bugs is $B(t)=\sum_{i=1}^L b_i(t)$. Finally, the number of defective parts, i.e.,
  those having at least one bug is $D(t)=\sum_{i=1}^L \Theta(b_i(t))$ \label{defD} where
  $\Theta(x)=1$ if $x>0$ and 0 otherwise, is the Heaviside function.

The dynamics of the program results from the interaction between those who use the program, detect bugs and complain about them, and those who try to correct them. To this end, we consider $N_u$ users. At each time step, each user is assumed to use one part, say, $i$, chosen at random, and to report a buggy behavior with a probability $P_u$ that depends on fraction of bugs $b_i/M$ in part $i$: this corresponds to the assumption that all the subparts are equally likely to be used in a time-step. Instead of assuming that $P_u$  is a generic function of $b_i/M$, we shall only retain its first order term:
  \be\label{pdetect} 
   P_u=\delta\,\frac{b_i}{M}. 
  \ee
The parameter $\delta$ describes the fraction of subparts used in a time-step, and also includes the propensity of the users to report bugs. Interestingly, keeping the zero-th order of $P_u$ is akin to suppose that the users report  bugs erroneously. Previous work assumed that $P_u=\textrm{cst}$.\cite{Botting} In our view, $P_u$ must contain a feedback from the actual number of bugs, otherwise bug reporting is a process completely disconnected from the actual program.


Each bug report only consists of the number of the buggy part because the user cannot describe in more details where the program is faulty. The bug list is hence a table indicating which parts are reportedly buggy, and its length is $R(t)$. It is the medium of interaction between the users and the programmers.

There are $N_p$ programmers. In addition to hunting bugs according to Eq.~\req{pdetect}, each of them tries to correct one part chosen at random from the bug report list, say, $i$, and reviews all the subparts of part $i$.  This process is assumed to fix a buggy subpart with probability $\phi$ and to break a working subpart with probability $\beta$;  Mathematically,
\be
P[s_{i,j}(t+1)=0|s_{i,j}(t)=1]=\phi,~~~~P[s_{i,j}(t+1)=1|s_{i,j}(t)=0]=\beta~~~~\forall j=1,\cdots,M.
\ee
The parameters $\phi$ and $\beta$ encode the programmers' abilities, which are chosen to be uniform; once again, this assumption does not change qualitatively the properties of the model; it is very simple to introduce heterogeneity here, at the cost of additional parameters. In the simplest version of our model, the programmers implement directly the modifications to the source code. In practice however, in larger projects, the programmers propose these modifications (so-called patches) to the maintainer.
  
The role of the maintainer is to determine whether a patch improves the code or not. The maintainer measures the number of bugs in the current code $\hat b_i$ and in the modified code $\hat b'_i$. He decides to accept the patch if he perceives that the patch is an improvement ($\hat b_i'<\hat b_i$), and removes part $i$ from the bug list. The
  measure is made as follows: he reviews the code of all sub-parts of part $i$, and detects correctly a buggy sub-part with probability $\nu$, and a working sub-part with probability $\omega$. 

How to estimate $\beta$, $\phi$, $\nu$ and $\omega$ from real data is discussed in section \ref{section:analytics}

This completes the definition of the model: the creation, detection and removal of bugs are fully specified, as are the interactions between the users, programmers, maintainer and the code. There is however a subtle point: the users are implicitely assumed to use always the latest version of the program. The time has come therefore to differentiate between open source and closed source projects. One of our aims is to show where the difference lies between these two strategies, and to this end, one is allowed for making it more pronounced by suitable assumptions. Bazaar open source projects release ``often and early'' new versions; running the latest, or a very recent, version of a given program is easy and probably quite common. Closed source projects on the other hand, do not release as nearly as often new versions, because they tend to prefer to release well-tested version whenever possible. The cost of new versions of commercial software also deters a fraction of the users to upgrade systematically. In summary, because of the very nature of these two development processes, {\em all other things being equal} OSS users necessarily upgrade faster than CSS users. This is translated in our model in the following way: OSS users always use the latest version, while CSS users upgrade every $T$ time steps. $T$ includes the lesser propensity of CSS to upgrade and the time between two releases; for the sake of simplicity, we shall speak of releases every $T$ time steps. The CSS users continue to report bugs of the latest release, while the CSS programmers work exclusively on the yet unreleased code. One objection to this hypothesis is that in real life there are alpha and beta testers, which greatly help the CSS programmers to hunt bugs. While this is correct, it is obvious that the number of alpha and beta testers is smaller than $N_u$, thereby reducing the ability of detecting bugs. In addition, we shall only study comparable situation, hence the {\em all other things being equal} mention. Nothing prevents the extension of our model to alpha/beta testers and to simulate various project configurations, but this is beyond the scope of the paper.

In closed-source projects, the programmers are faced with a dilemma
when a part is reported as buggy by a user after it has been already
tentatively fixed since the last release. Indeed, the users report bugs on the
last release, whereas the programmers work on the next one, both
gradually diverging. The programmers can either ignore bug reports on
an already modified part, or modify again the current code. In the
latter case, the modification can be systematic, or after verification
that the part in the current code is also seemingly buggy (according to
Eq.~\req{pdetect}). Without verification, $D(t)$ is not a monotonically
decreasing quantity, as a newly bug-free part can be partly broken by this
process.

Finally all the parameters of this model can be changed at each time step, and will be in Section 4, thus

\section{Results}
\noindent

We shall first report numerical experiments and then propose qualitative analytical explanations of the observed behaviours. Our aim is not to validate our with real-life figures, but to check whether it behaves reasonably, and what parameters are the most important for improving the quality of a software project.

\begin{figure}
\centerline{\includegraphics[height=5cm]{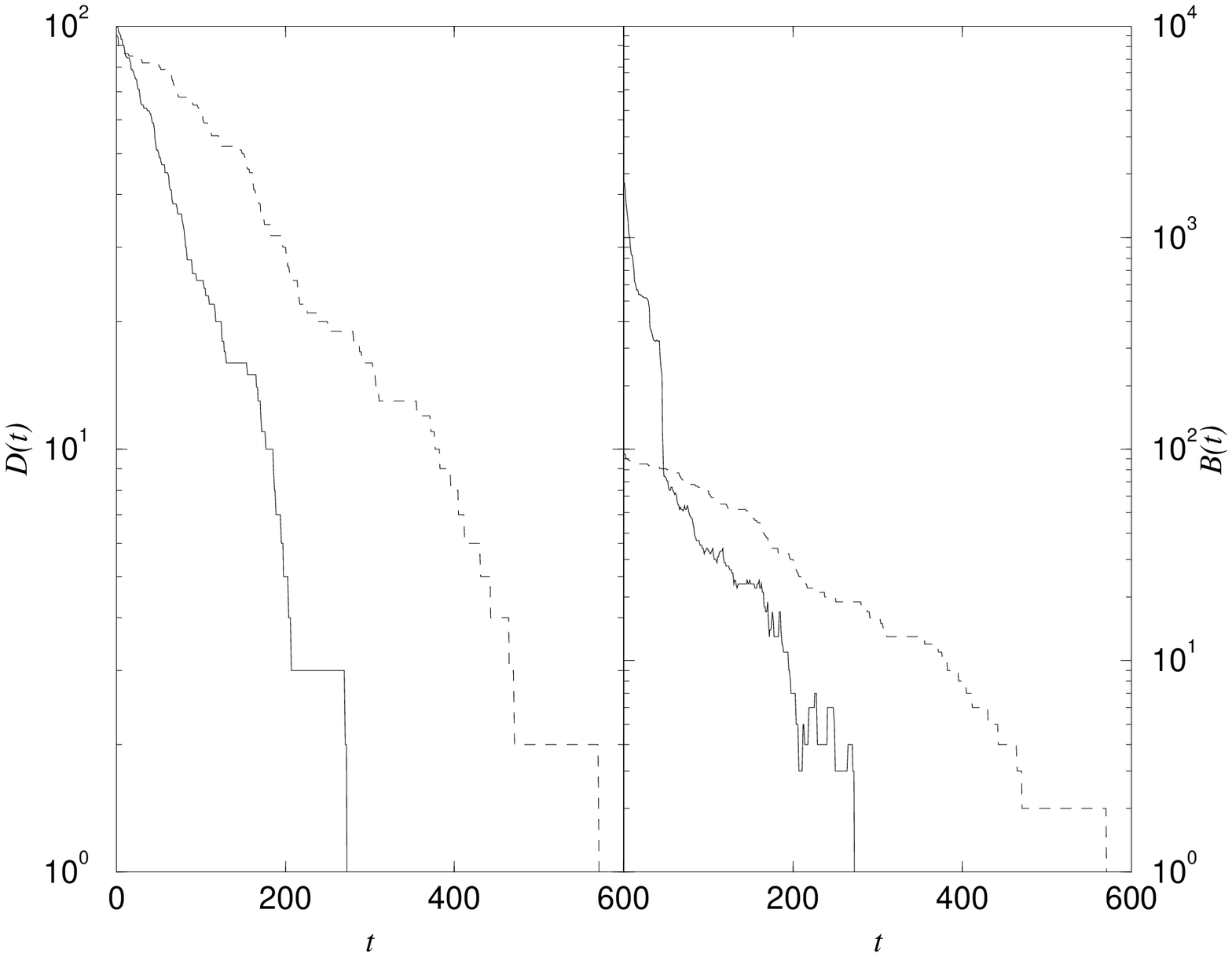}}
\fcaption{Number of defective parts (left panel) and  number of
  bugs (right panel) in an open source project (continuous lines),
  closed source projects with no bug report resubmission allowed
  between releases (dashed lines) $L=100$, $M=20$, $N_u=100$,
  $N_p=10$, $\delta=1$, $\phi=0.9$, $\beta=0.1$, $\omega=0.9$,
  $\nu=0.9$. $T=1$ for OSS and $T=50$ for CSS.}
\label{runBOTH}
\end{figure}

Fig. \ref{runBOTH} reports the typical dynamical behavior of
projects that start from scratch ($s_{i,j}=1$ for all $i$
and $j$): the number of defective parts $D(t)\propto \exp(-\lambda t)$
for large $t$, as often assumed in reliability growth models.\cite{SoftwareEng}  At a
more microscopic level, one can distinguish two phases: in the early
easy stage, the users find and report many bugs, keeping $R(t)\gg
N_p$. The vast majority of bugs are fixed during this phase, where
$B(t)$ decreases linearly with time. When $R(t)\sim N_p$, a slow
regime appears, where $B(t)\sim \exp(-t/\tau)$; in this regime, the
average number of bugs per defective part $B(t)/D(t)$ is small and
fluctuates around a value that depends on the chosen parameters.

Ignoring bug reports on already modified
code is the best option for CSS; this even
outperforms OSS at short time scales, because the programmers
only work on fully buggy parts, hence the bug fixing rate is higher
(right panel of Fig. \ref{runBOTH}). Verification is generally a bad
idea when bugs are sparse, because the probability that both
a user and a programmer agree that a part is buggy is small, hence
verification slows down the process.  
\begin{figure}
\centerline{\includegraphics[height=5cm]{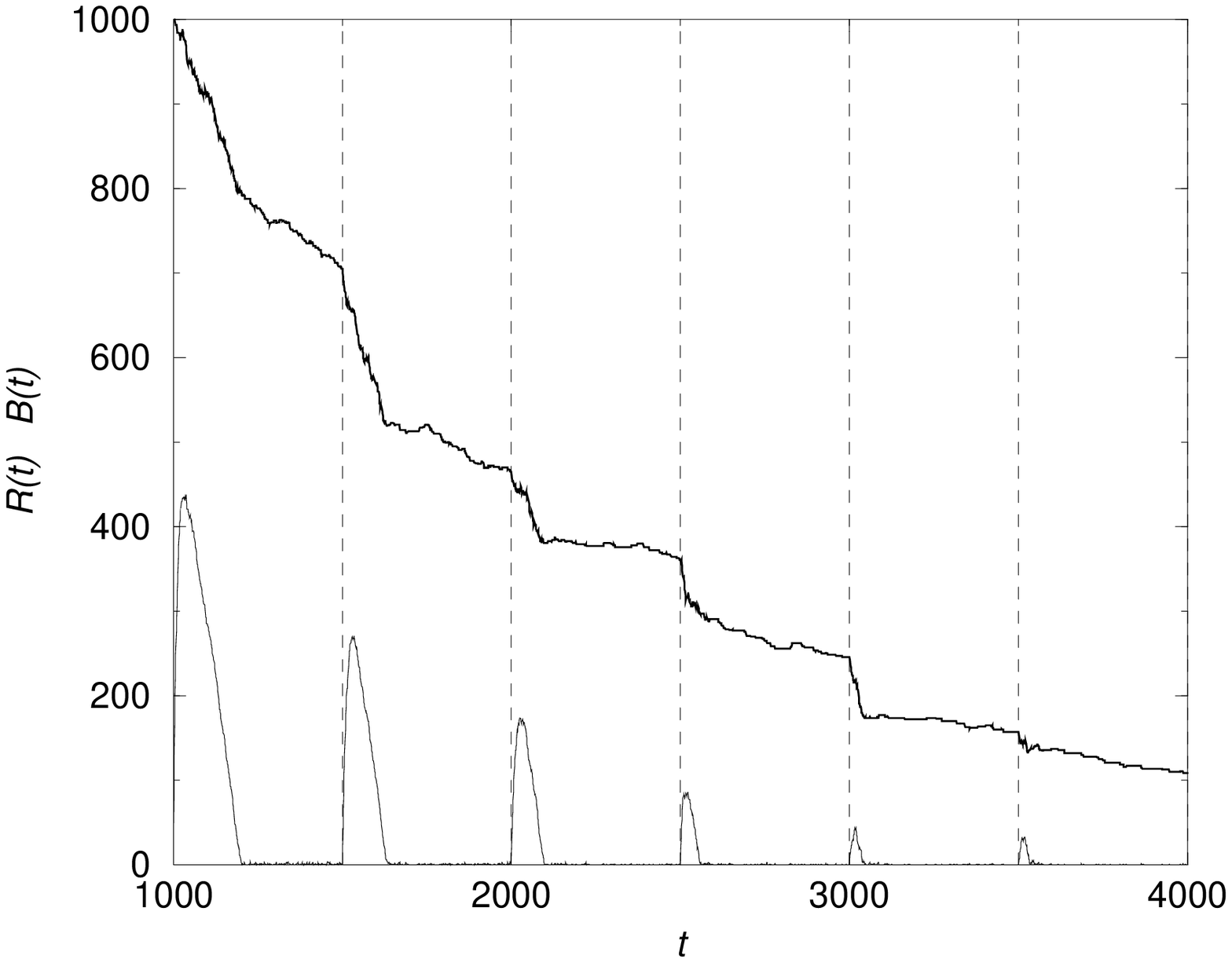}}
\fcaption{Dynamics of the number of bugs and bug reports in closed
  source projects ($T=500$).
  $L=1000$, $M=40$, $N_u=1000$, $N_p=10$, $\delta=1$, $\phi=0.9$, $\beta=0.1$, $\omega=0.9$,
  $\nu=0.9$}
\label{RBt}
\end{figure}

The global temporal evolution
can be characterized by the time to completion, i.e., the time needed
to obtain a bug-free project, denoted by $t_c$. We shall investigate
in particular its average $\avg{t_c}$ over several runs. For the sake
of speed, we stop the simulations when $B(t)=1$: since the decrease of $B(t)$ is exponential in this phase, $\avg{t_c}$ would be roughly doubled if one waited until $B=0$.

\begin{figure}
\centerline{\includegraphics[height=5cm]{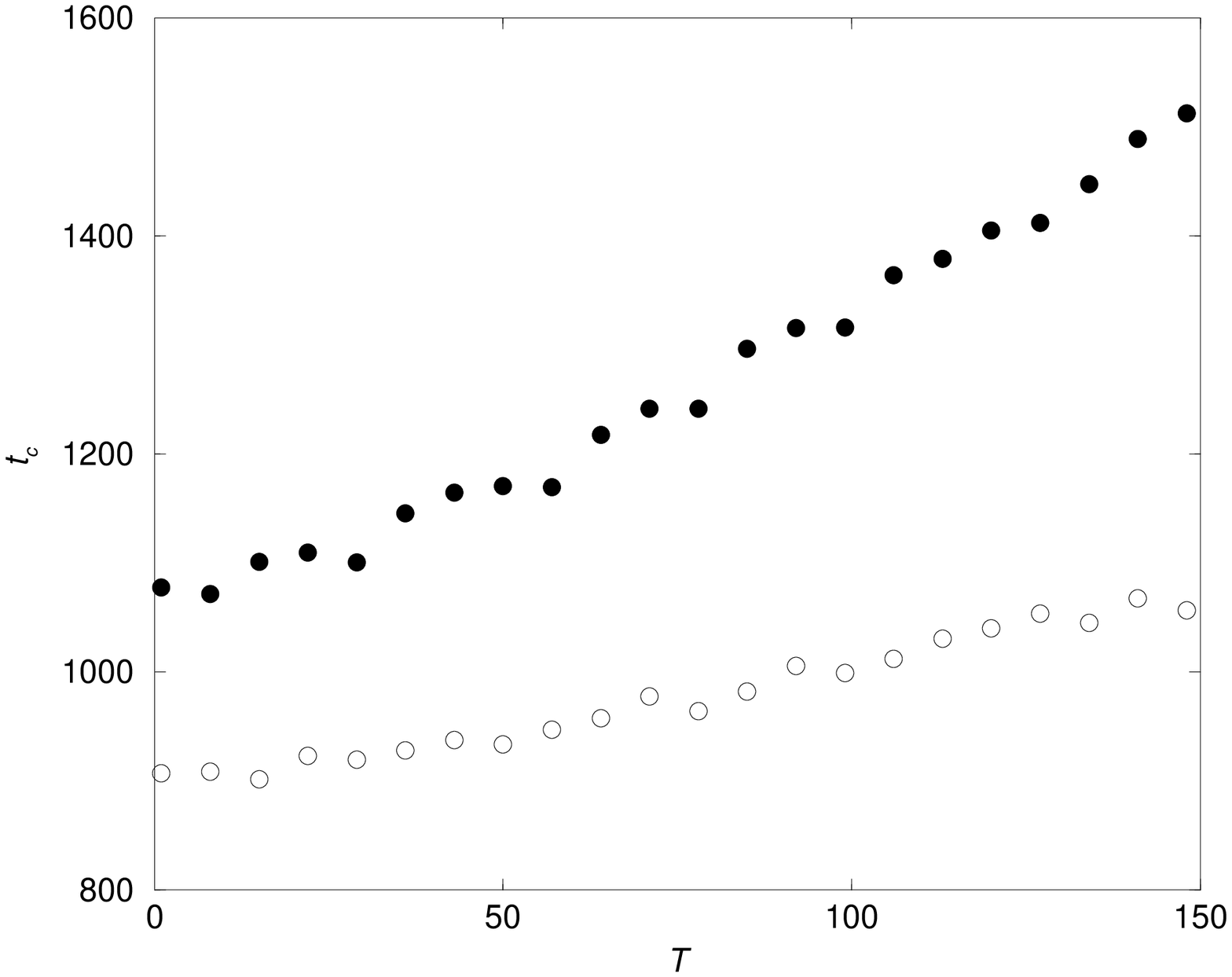}}
\fcaption{Average time to completion $t_c$ as a function of the time between releases $T$.  Full
  symbols: $L=1000$, $M=10$, empty symbols $L=2000$, $M=5$; $N_u=100$,
  $N_p=10$, $\delta=1$, $\phi=0.9$, $\beta=0.1$, $\omega=0.9$,
  $\nu=0.9$ average over 100 runs.}
\label{tauT}
\end{figure}
\begin{figure}
\centerline{\includegraphics[height=5cm]{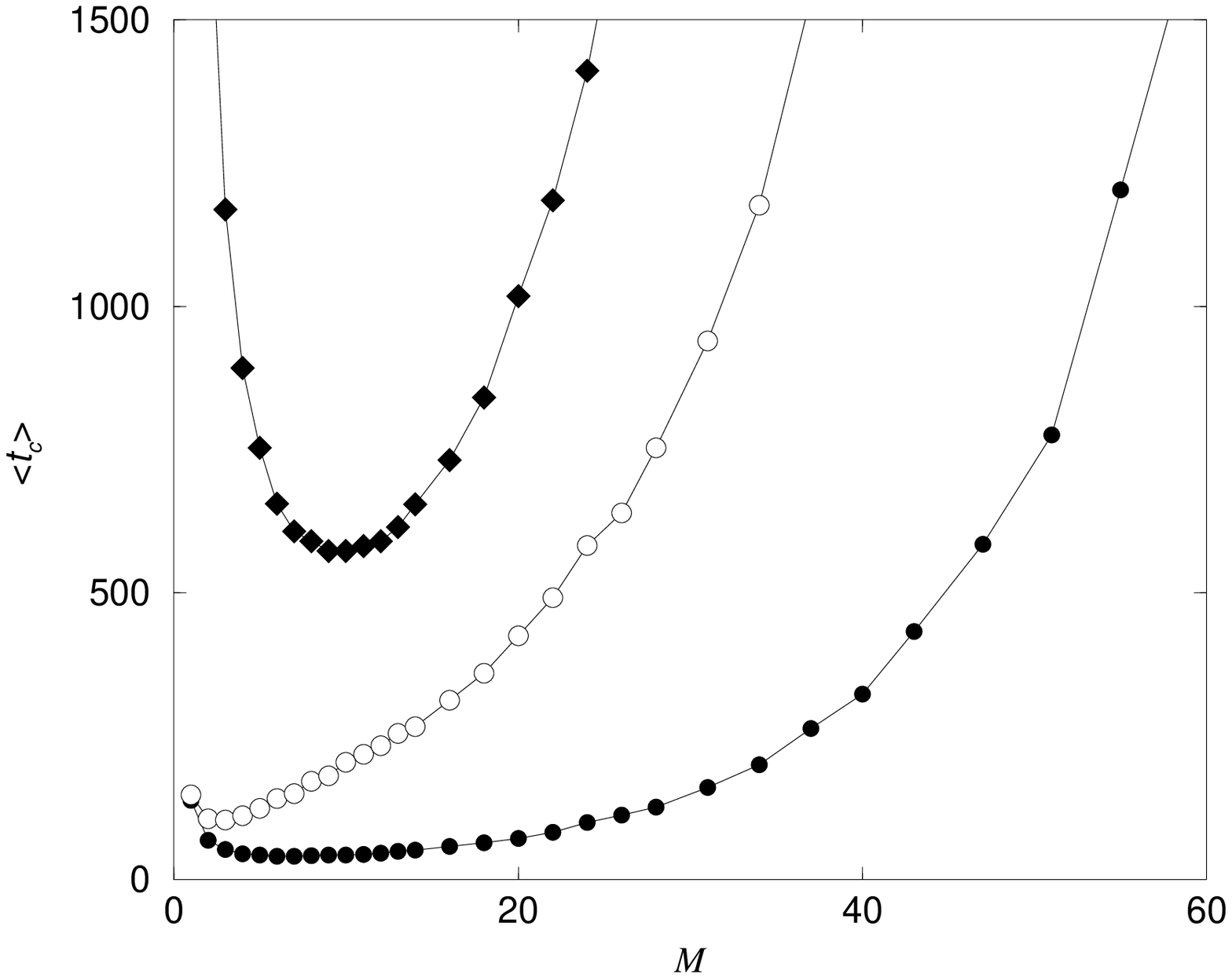}}
\fcaption{Average time to completion $\avg{t_c}$ as a function of the number of
  subparts per part $M$ at constant size $LM=1000$ for open source
  projects (full circles), closed source without bug report
  resubmission between two releases (empty squares), and closed source
  with bug report resubmission (full diamonds).
  $N_u=1000$, $N_p=10$, $\delta=1$, $\phi=0.9$, $\beta=0.1$, $\omega=0.9$,
  $\nu=0.9$, closed source $T=150$, average over 100 runs. Continuous
  lines are for eye-guidance only.}
\label{Mtc}
\end{figure}
The average time to completion increases roughly linearly with $T$
(Fig~\ref{tauT}), hence, closed source projects are always slower to
reach a perfect state than open source projects, {\em all other parameters being equal}. The reason why
increasing $T$ penalizes the performance of the project is 
obvious from Fig~\ref{RBt}: after each release, the number of
relevant bug reports coming from the users falls rapidly to zero, and
the programmers are left on their own; as a consequence, the fast and slow
regimes alternate.

Suppose that the
project size $S$ is known in advance and fixed to $S=LM$. What $L$ and
$M$ are optimal? Fig. \ref{Mtc} plots $\avg{t_c}$ as a function of $M$ at
fixed $S=LM$ for open and closed source. It turns out that there is
always an optimal value of $M$ for any set of parameters or project
type, whose position depends on all the parameters, and that the plot is much shallower for OSS projects, implying that the values of $L$ and $M$ are not crucially important for them..

\begin{figure}
\centerline{\includegraphics[height=5cm]{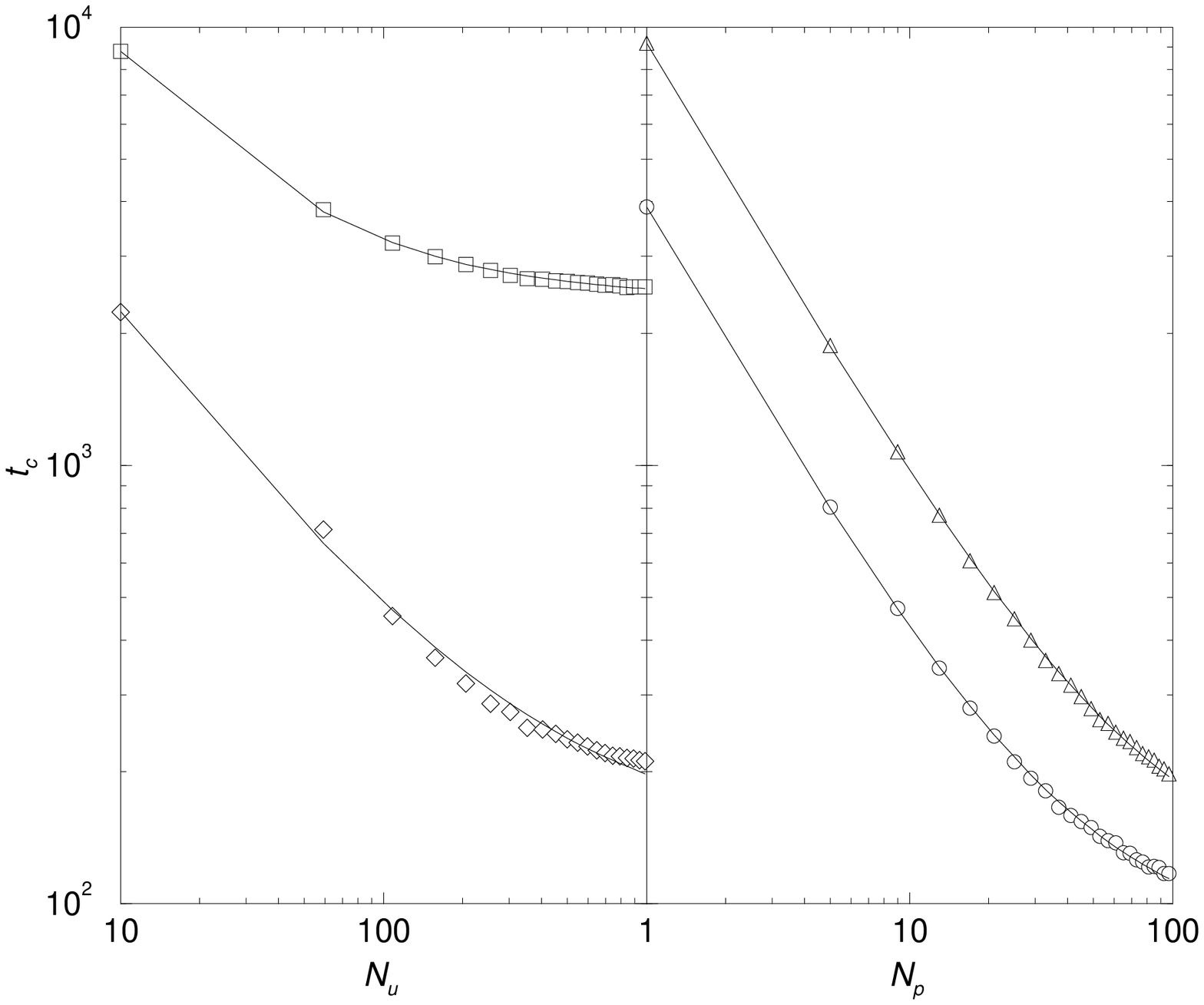}}
\fcaption{Average time to completion $t_c$ versus the number of users
  (left panel) and the number of programmers (right panel).
   $L=1000$, $M=10$, $\delta=1$, $\omega=0.9$,
  $\nu=0.9$; $N_p=10$ (left panel), $N_u=1000$ (right
   panel). $\phi=0.9$, $\beta=0.1$ (diamonds and circles), $\phi=0.7$,
   $\beta=0.3$ (squares), and $\phi=0.6$, $\beta=0.1$ (triangles); average over 100 runs.}
\label{tcNuNp}
\end{figure}

Fig. \ref{tcNuNp} shows how $\avg{t_c}$ depends on $N_u$ and $N_p$:
$\avg{t_c}$ is reasonably well fitted by $c_1+c_2N_u^{-\alpha}$ with
$\alpha\sim 1$: the rate of improvement is slow as $N_u$ increases;
even worse, it reaches a plateau $c_1$ whose value depends on the
number of programmers $N_p$ and their abilities; the exponent $\alpha$
also depends on the programmers abilities, but not on their number.
Similarly, adding more programmers decreases $\avg{t_c}$.  In this
case, the exponent $\alpha$ depends on the number of users, but not on
the programmers abilities. In other words, hiring more programmers or
having more users is an inefficient way of improving the speed of
debugging when $N_u$ or $N_p$ is large enough.
 Interestingly, having better programmers decreases $\avg{t_c}$, while the abilities of the
maintainer has much less dramatic an influence. 

From our model we conclude that bazaar OSS methodology has the shortest average time to completion, all other parameters being equal, which is precisely the argument of Raymond.\cite{ESR} However, this does
not mean that bazaar OSS is always faster: cathedral OSS or CSS
projects with a better set of parameters (more programmers, better
programmers, more users) can outperform bazaar OSS even with
large time between releases. On the other hand, the quality of bazaar
OSS programmers does not need to be as high as those
of cathedral OSS or CSS projects in order to achieve the same time of
convergence to the bug-free state. Finally, our model suggests that cathedral OSS and 
CSS projects should try to minimize $T$ so as to decrease
their convergence time, for instance by implementing automatic upgrades. This of course requires that the users do not perceive a relatively high upgrade rate as an indication of low-quality software.

\begin{figure}
\centerline{\includegraphics[height=5cm]{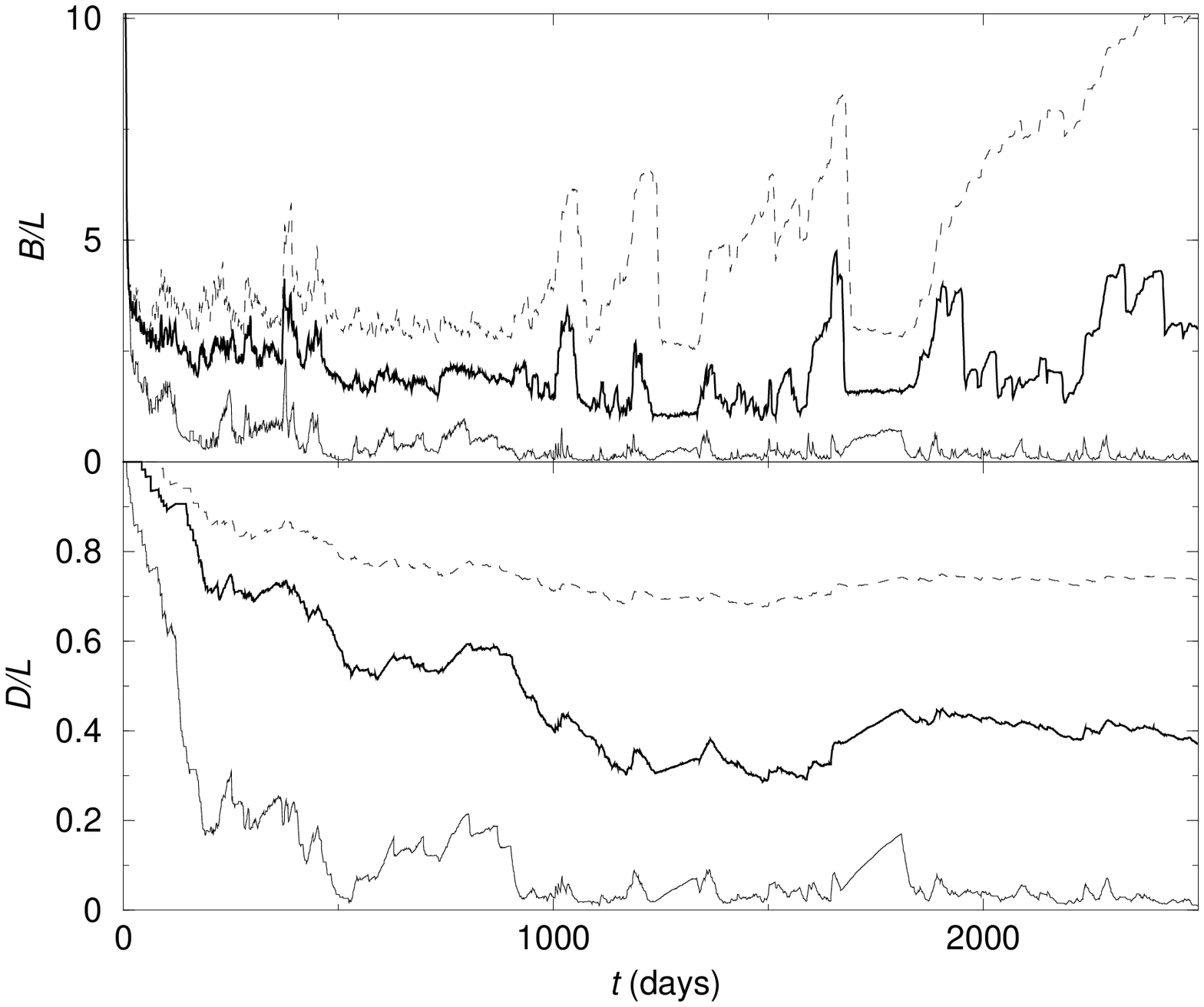}}
\fcaption{Average number of bugs per part (upper panel) and fraction of
  defective parts  (lower panel) versus time in a reconstructed Linux
  history. $M=20$, $\delta=1$, $\phi=0.8$, $\beta=0.05$
  (thin line) $\beta=0.1$ (thick lines), $\beta=0.2$ (dashed line), $\omega=0.5$,
  $\nu=0.5$}
\label{Linuxev}
\end{figure}

\section{Mathematical analysis}
\label{section:analytics}
While it is not possible to solve this model exactly, some understanding can be obtained through simple analysis. Bug reports are filled with an average frequency of $(N_u+N_p)\frac{\delta B(t)}{L M}$: each user/programmer has a probability $D(t)/L$ to use a  part that contains at least one bug (by definition of $D(t)$, see page \pageref{defD}), and a probability $\delta\frac{B(t)}{D(t)}/M$ to report a bug. The probability that exactly two users/programmers use the same part and fill bug reports is proportional to $(N_u+Np)(N_u+N_p-1)(\delta D/L)^2\avg{(b_i/M)^2}_D/2$ where $\avg{}_D$ stands for the average over all the defective parts; in later stages of debugging, this probability may be very small for appropriate parameters, and redundant bug reporting can be neglected. 

The probabilily that a reported bug is already in the bug list is roughly $R(t)/D(t)$ (this assumes that all the defective parts have about the same number of bugs). Therefore, the number of relevant bug reports submitted at each time step is about $(N_u+N_p)\frac{\delta B(t)}{L M}(1-R(t)/D(t))$ on average (without taking into account redundant report filling). Finally, neglecting the role of the maintainer, i.e. always accepting a patch, the number of bug reports removed from the list at each time step is $\min[R(t),N_p]$, hence
\be
R(t+1)=R(t)+(N_u+N_p)\frac{\delta B(t)}{L M}(1-R(t)/D(t))-\min[R(t),N_p]+n_R(t)
\ee
where $n_R(t)$ is a white Gaussian noise of zero average.

The dynamics of the programmers is relatively simple to analyse: assume that part $i$ is in the report list and is picked up by a programmer at time $t$. Then (dropping the index)
\be\label{eq:b(t)}
b(t+1)=b(t)\phi+[M-b(t)]\beta+n(t)
\ee
where $n(t)$ is a noise term of zero average and $\avg{n(t)n(t')}=\delta_{t,t'}[b(t)\phi(1-\phi)+(M-b(t))\beta(1-\beta)]$ where $\avg{x}$ stands for the average of $x$ over time.
\be
b^*=\frac{\beta M}{1+\beta-\phi}
\ee
which is reached exponentially fast
A bug-free state corresponds to $b=0$.  When the users do not report imaginary bugs, that is, when $P_u(b/M)$ does not contain a constant, the boundary $b=0$ is absorbing: if there are no more bugs in this part, no one will ever be found in it. 

We thus must conclude that the only reason why a bug-free state can be reached when the programmers sometimes break working code ($\beta>0$) is via fluctuations. The fluctuations around $b^*$ are of order $\sqrt{M}$ but $b^*$ is of order $M$, hence, the larger $M$, the more difficult it is to reach $b=0$ by chance, which explains why the time to completion increases as $M$ increases in Fig. \ref{Mtc}. Assuming that $M$ is constant and no bias ($\phi=\beta$), it is easy to solve a diffusion equation in the interval $[0,M]$ with reflecting boundary at $b=M$ and absorbing boundary at $b=0$ with initial condition $b(0)=M$, which gives 
\be
P(b,t)=\sqrt{\frac{2}{M}}\sum_{n\ge 1}\sin\cro{\frac{\pi}{M}\pr{\frac{1}{2}+2n}b}e^{-C\cro{\frac{\pi}{M}\pr{\frac{1}{2}+2n}}^2t}
\ee
where the diffusion constant $C\propto M$. For large times, the survival probability $S(t)=\sum_b P(b,t)$ decreases as its slowest component, hence
\be
S(t)\sim \sqrt{M}e^{-\frac{C^2\pi^2}{4M}t}
\ee
This shows that $t_c\sim M^{3/2}$ for large $M$ in this approximation. Given the fact that we did not take into account the potential that attracts $b$ around $b^*$, this is clearly an under-estimation. 

Since the number of bug fluctuates around $b^*$, it is sensible to assume that every part with at least one bug has $b^*$ bugs and that it is the number of buggy parts that decrease.

The role of the maintainer is difficult to describe analytically as the formulae are cumbersome. But simple approximations bring some light. The average number of bugs $d(b)$ detected by the maintainer in a part that contains $b$ buggy sub-parts is $b\nu+(M-b)(1-\omega)$, with fluctuations $\sigma^2(b)=b\nu(1-\nu)+(M-b)\omega(1-\omega)$. Therefore, approximating the probability that the maintainer perceives $\hat b$ bugs when there are $b$ bugs by a Gaussian distribution of average $d(b)$ and variance $\sigma(b)$, the probability that a maintainer rejects a patch with $b'$ buggy sub-parts if the current code has $b$ buggy sub-parts is, in a first approximation, 
\be\label{eq:Pbhat}
P(\hat b'>\hat b)\simeq\frac{1}{2}+\int_{-\infty}^{\infty}\frac{e^{-\frac{(x-d(b))^2}{2\sigma(b)^2}}}{\sqrt{2\pi}\sigma(b)}\frac{1}{2}\erf\pr{\frac{x-d(b')}{\sqrt{2}\sigma(b')}}\dd x.
\ee
$d(b)$ is an increasing function of $b$ if $\nu+\omega>1$, i.e., if the maintainer is sufficiently gifted. In that case, it is easy to convince oneself that $P(\hat b'>\hat b)>\frac{1}{2}$ if $b'>b$. If $\nu+\omega<1$, the fewer the bugs, the more bugs the maintainer thinks there are and $P(\hat b'>\hat b)<\frac{1}{2}$ when $b'>b$. The special case $\nu+\omega=1$ yields simply $P(\hat b'>\hat b)=\frac{1}{2}$: he tosses a coin in order to determine whether to commit patches or not. In any case, the maintainer's role is merely that of a timescale: depending on his abilities, he will delay or speed up the rate of acceptance of good patches, and his quality can be defined as $P(\hat b'<\hat b|b'<b)$.

How to measure $\beta$, $\phi$, $\omega$ and $\nu$ in real data can be done by noting that $b_i$ can be observed indirectly via the rates of bug reports, which are proportional to the density of faulty sub-parts. Dividing the evolution equation of $b_i$ \req{eq:b(t)} by $M$, one obtains the evolution of the density of bugs $\rho_i$, i.e. the bug reporting rate for part $i$ if $\delta=1$. Estimating $\beta$ and $\phi$ is done by computing the average and the fluctuations of all $\rho_i(t)$, conditional on $\rho_i(t)$. Since both of them are proportional to $\rho_i(t)$, one can simply plot these two quantities versus $\rho_i$ for all $i$, and perform least squares linear fits.
Once $\beta$ and $\phi$ are known, the parameters of the maintainer can be estimated in a similar way. First one should replace all references to bug numbers $b$ and $\hat b$ by rates $\rho$ and $\hat \rho=\hat b/M$, etc,  in Eq. \req{eq:Pbhat}. Then since $\beta$ and $\phi$ are known, one also knows $P(\rho'|\rho)$, and, for any given estimation of $\omega$ and $\nu$, one can therefore compute $P(\hat \rho'>\hat \rho)$, which allows for the measure of the maintainer's characteristics.

\section{Dynamics of Linux}\label{Linux}
\noindent

As shown above, the quality of OSS programmers does
not need to be as high as CSS ones in order to achieve a bug-free
state as rapidly {\em all other things being equal}. However, this does not mean that it can be vanishingly small:
the quality of OSS programmers in successful projects has a lower bound. As an illustration, let us consider the history of Linux. From version 1.0, the number of programmers
$N_p(t)$ can be obtained from the CREDITS file. It is well fitted by a
$80+0.1d$ where $d$ is the number of days since Linux 1.0.  The number
of users $N_u(t)$ is hard to estimate because of the free nature of
Linux. Four estimates available on Internet\cite{RH} can be fitted with with a
power law $N_u(y)\propto [y-1991]^{3.6}$ where $y$ denotes the year
and 1991 is Linux' date of birth.  The size $S(t)$ can be measured in
number of lines divided by the typical number of lines in a subpart;
it has been fitted with a quadratic function,\cite{Godfrey} which is
consistent with the linear increase of $N_p(t)$, as
$S(t+1)-S(t)\propto N_p(t)$. 
We can translate $S(t)$ (measured directly in the
source code) into $L(t)$ and $M$ by supposing that each subpart
contains $M=20$ lines of code. The other parameters that cannot be determined
directly from Linux are obviously the qualities of the programmers and
the maintainers ($\delta$, $\beta$, $\phi$, $\omega$ and $\nu$).  In an attempt to be pessimistic, and without prejudice for Linus Tovarlds, we considered a random
maintainer, that is, $\omega=\nu=1/2$, and sub-optimal $M$;
$\delta$ is fixed to $1$.  The new parts of Linux are assumed to be
first completely buggy.

Figure \ref{Linuxev} shows a transition
between two very different behaviors depending on the choice of $\phi$
and $\beta$: if the programmers abilities are high enough, Linux
converges fast to the slow regime, which is stable with respect to
sudden increases of the system size, and where
the number of bug per part decreases as a function time (e.g. $\phi=0.8$ and
$\beta=0.05$); if the quality of the programmers is too low, Linux
falls into the region where the number of bugs makes large excursions
(e.g.  $\phi=0.8$, $\beta=0.15$), resulting in a dramatic decrease of
reliability. Since Linux is known to be stable, this
shows that the quality of Linux programmers has a lower bound.\footnote{Note that we do not pretend to measure the real quality of Linux's programmers.}  Therefore, super-linear software growth can be
sustained if the programmers are sufficiently skilled, and if there
are enough programmers. The above
picture can be generalized to any project whose
parameters evolve in time. 

\section{Outlook}
\noindent

Our model is designed to be simple and generic so as to provides a generic modeling 
framework. Every simplifying assumption can be remedied. For instance, relevant extensions include users that upgrade their program after some delay, for instance, only after they have found a bug, or randomly after a new release has been released. Heterogeneous rates ($\delta$, $\beta$, $\phi$, $\omega$, $\nu$) and part sizes should be drawn from a suitable distribution; this will result in non-linearities in Eq.~\ref{pdetect}. One could also impose a restriction on the number of subparts that a programmer is able to review in one time-step. An important
assumption of the present model is the independence of the parts, whereas
they are linked by a scale-free asymmetric network,\cite{Sole,SoftLinks} hence, bugs can propagate on this graph and affect other parts, making debugging harder.\cite{SoftLinks} The next step is therefore to study this model on
scale-free networks. In addition, the number of modifications per programmer is a truncated
power-law in Linux, as is the number of bugs assigned and corrected per programmer in Mozilla.\cite{Mozilla,unpubl} Therefore it may be possible that the decay of the number of bugs will not be exponential anymore, but follow a power-law, as assumed in some reliability growth models.\cite{Duane,RGM} Assigning a higher
or lower bug fixing priority to the parts that have more bug reports
may interact with the emergence of power-laws in the decrease of the
number of bugs. Finally, programmers and maintainer could be modelled as agents that could learn from their mistakes. All these modifications are likely to reveal many fascinating
subtleties of bug dynamics and the relationship between the micro- and macroscopic levels

\section{Acknowledgments}

DC thanks Wadham College for support. Useful suggestions from Ani Calinescu and Matteo Marsili are gratefully acknowledged.


\begin{thebibliography}{1}
\bibitem{SoftwareEng} J. Sommerville, {\em Software Engineering},
  Addison-Wesley, Harlow (2001)
\bibitem{Agile}  R. C. Martin, {\em Agile Software Development,
    Principles, Patterns, and Practices }, Prentice Hall (2002)
\bibitem{RGM} A. Biroli, {\em Reliability Engineering, Theory and Practice}, Springer,
  Berlin (1999)
\bibitem{Bishop} Bishop and Blomfield, A conservative theory for long-term reliability growth prediction, IEEE Trans. Rel. {bf 45}(4), 550 (1996)
\bibitem{RossBall} R. M. Brady, R. J. Anderson, R. C. Ball, Murphy's law, the fitness of evolving species, and the limits of software reliability, Cambridge Technical Report UCAM-CL-TR-471 (1999)
\bibitem{Lyu} Software Reliability Engineering Handbook, M. R. Lyu ed, McGraw-Hill Education (1995)
\bibitem{web} \url{http://irb.cs.uni-magdeburg.de/sw-eng/us/bibliography/bib_main.shtml}
\bibitem{Botting} R. Botting, Some Stochastic Models of Software
  Evolution, paper presented at Systemics, Cybernetics and
  Informatics'2002, SCI {\bf 1}, Orlando, USA, July 2002 
\bibitem{fsf} Free Software Foundation {\tt www.fsf.org} 
\bibitem{linux} Linux {\tt www.kernel.org}
\bibitem{apache} Apache {\tt www.apache.org}
\bibitem{ESR} E. S. Raymond, {\em The Cathedral \& the Bazaar},
  O'Reilly (2001), (available at {\tt catb.org/~esr/writings/cathedral-bazaar)}
\bibitem{RH} RedHat {\tt www.redhat.com/about/corporate/milestones.html}
\bibitem{Godfrey} M. W. Godfrey and Q. Tu, Evolution in Open Source
  Software: A Case Study, Proc. of the 2000 Intl. Conference on Software Maintenance (ICSM-00), San Jose, California, October 2000
\bibitem{Mozilla} Mozilla {\tt www.mozilla.org}
\bibitem{Sole} S. Valverde, R. Ferrer, R.  V. Sole, Scale free networks from optimal design, 
Europhysics Letters {\bf 60}, 512-17 (2002)
\bibitem{SoftLinks} D. Challet and A. Lombardoni, Bug propagation and debugging in asymmetric software structures, Phys. Rev. E {\bf 70}, 046109 (2004).
\bibitem{Duane} J. T. Duane, Learning curve approach to reliability, monitoring, 
IEEE Transactions on Aerospace, AS-2, {\bf 2}, 563-566 (1964).
\bibitem{unpubl} D. Challet, Open-source software development: the role of contributors, unpublished (2003), material available on request
\end{thebibliography}
\end{document}